
%
\def\unlockat{\catcode`\@=11}
\def\lockat{\catcode`\@=12}
\unlockat
\def\d@f@ult{} \newif\ifamsfonts \newif\ifafour
\def\m@ssage{\immediate\write16}  \m@ssage{}
\m@ssage{hep-th preprint macros.  Last modified 16/10/92 (jmf).}
\message{These macros work with AMS Fonts 2.1 (available via ftp from}
\message{e-math.ams.com).  If you have them simply hit "return"; if}
\message{you don't, type "n" now: }
\endlinechar=-1  
\read-1 to\@nswer
\endlinechar=13
\ifx\@nswer\d@f@ult\amsfontstrue
    \m@ssage{(Will load AMS fonts.)}
\else\amsfontsfalse\m@ssage{(Won't load AMS fonts.)}\fi
\message{The default papersize is A4.  If you use US 8.5" x 11"}
\message{type an "a" now, else just hit "return": }
\endlinechar=-1  
\read-1 to\@nswer
\endlinechar=13
\ifx\@nswer\d@f@ult\afourtrue
    \m@ssage{(Using A4 paper.)}
\else\afourfalse\m@ssage{(Using US 8.5" x 11".)}\fi
\nonstopmode
%
%

\font\twelverm=cmr12
\font\ninerm=cmr9
\font\sixrm=cmr6
\font\fourteenbf=cmbx12 scaled\magstep1
\font\twelvebf=cmbx12
\font\ninebf=cmbx9
\font\sixbf=cmbx6
\font\fourteeni=cmmi12 scaled\magstep1      \skewchar\fourteeni='177
\font\twelvei=cmmi12                        \skewchar\twelvei='177
\font\ninei=cmmi9                           \skewchar\ninei='177
\font\sixi=cmmi6                            \skewchar\sixi='177
\font\fourteensy=cmsy10 scaled\magstep2     \skewchar\fourteensy='60
\font\twelvesy=cmsy10 scaled\magstep1       \skewchar\twelvesy='60
\font\ninesy=cmsy9                          \skewchar\ninesy='60
\font\sixsy=cmsy6                           \skewchar\sixsy='60
\font\fourteenex=cmex10 scaled\magstep2
\font\twelveex=cmex10 scaled\magstep1

\ifamsfonts
   \font\ninex=cmex9
   
   \font\sixex=cmex7 at 6pt
   
\else
   \font\ninex=cmex10 at 9pt
   
   \font\sixex=cmex10 at 6pt
   
\fi
\font\fourteensl=cmsl10 scaled\magstep2
\font\twelvesl=cmsl10 scaled\magstep1

\font\sevensl=cmsl10 at 7pt
\font\sixsl=cmsl10 at 6pt

\font\fourteenit=cmti12 scaled\magstep1
\font\twelveit=cmti12

\font\fourteentt=cmtt12 scaled\magstep1
\font\twelvett=cmtt12
\font\fourteencp=cmcsc10 scaled\magstep2
\font\twelvecp=cmcsc10 scaled\magstep1

\ifamsfonts
   
\else
   
\fi
\newfam\cpfam
\font\fourteenss=cmss12 scaled\magstep1
\font\twelvess=cmss12
\font\tenss=cmss10
\font\niness=cmss9

\font\sevenss=cmss8 at 7pt
\font\sixss=cmss8 at 6pt
\newfam\ssfam
\newfam\msafam \newfam\msbfam \newfam\eufam
\ifamsfonts
 \font\fourteenmsa=msam10 scaled\magstep2
 \font\twelvemsa=msam10 scaled\magstep1
 \font\tenmsa=msam10
 \font\ninemsa=msam9
 \font\sevenmsa=msam7
 \font\sixmsa=msam6
 \font\fourteenmsb=msbm10 scaled\magstep2
 \font\twelvemsb=msbm10 scaled\magstep1
 \font\tenmsb=msbm10
 \font\ninemsb=msbm9
 \font\sevenmsb=msbm7
 \font\sixmsb=msbm6
 \font\fourteeneu=eufm10 scaled\magstep2
 \font\twelveeu=eufm10 scaled\magstep1
 \font\teneu=eufm10
 \font\nineeu=eufm9
 
 \font\seveneu=eufm7
 \font\sixeu=eufm6
 \def\hexnumber@#1{\ifnum#1<10 \number#1\else
  \ifnum#1=10 A\else\ifnum#1=11 B\else\ifnum#1=12 C\else
  \ifnum#1=13 D\else\ifnum#1=14 E\else\ifnum#1=15 F\fi\fi\fi\fi\fi\fi\fi}
 \def\hexmsa{\hexnumber@\msafam}
 \def\hexmsb{\hexnumber@\msbfam} 
\fi
\newdimen\b@gheight             \b@gheight=12pt
\newcount\f@ntkey               \f@ntkey=0
\def\f@m{\afterassignment\samef@nt\f@ntkey=}
\def\samef@nt{\fam=\f@ntkey \the\textfont\f@ntkey\relax}
\def\rm{\f@m0 }
\def\mit{\f@m1 }
\def\cal{\f@m2 }
\def\it{\f@m\itfam}
\def\sl{\f@m\slfam}
\def\bf{\f@m\bffam}
\def\tt{\f@m\ttfam}
\def\caps{\f@m\cpfam}
\def\ssf{\f@m\ssfam}
\ifamsfonts
 \def\msa{\f@m\msafam}
 \def\msb{\f@m\msbfam} \let\bb=\msb
 \def\eu{\f@m\eufam}
\else
 \let \bb=\bf \let\eu=\bf
\fi
\def\fourteenpoint{\relax
    \textfont0=\fourteencp          \scriptfont0=\tenrm
      \scriptscriptfont0=\sevenrm
    \textfont1=\fourteeni           \scriptfont1=\teni
      \scriptscriptfont1=\seveni
    \textfont2=\fourteensy          \scriptfont2=\tensy
      \scriptscriptfont2=\sevensy
    \textfont3=\fourteenex          \scriptfont3=\twelveex
      \scriptscriptfont3=\tenex
    \textfont\itfam=\fourteenit     \scriptfont\itfam=\tenit
    \textfont\slfam=\fourteensl     \scriptfont\slfam=\tensl
      \scriptscriptfont\slfam=\sevensl
    \textfont\bffam=\fourteenbf     \scriptfont\bffam=\tenbf
      \scriptscriptfont\bffam=\sevenbf
    \textfont\ttfam=\fourteentt
    \textfont\cpfam=\fourteencp
    \textfont\ssfam=\fourteenss     \scriptfont\ssfam=\tenss
      \scriptscriptfont\ssfam=\sevenss
    \ifamsfonts
       \textfont\msafam=\fourteenmsa     \scriptfont\msafam=\tenmsa
         \scriptscriptfont\msafam=\sevenmsa
       \textfont\msbfam=\fourteenmsb     \scriptfont\msbfam=\tenmsb
         \scriptscriptfont\msbfam=\sevenmsb
       \textfont\eufam=\fourteeneu     \scriptfont\eufam=\teneu
         \scriptscriptfont\eufam=\seveneu \fi
    \samef@nt
    \b@gheight=14pt
    \setbox\strutbox=\hbox{\vrule height 0.85\b@gheight
                                depth 0.35\b@gheight width\z@ }}
\def\twelvepoint{\relax
    \textfont0=\twelverm          \scriptfont0=\ninerm
      \scriptscriptfont0=\sixrm
    \textfont1=\twelvei           \scriptfont1=\ninei
      \scriptscriptfont1=\sixi
    \textfont2=\twelvesy           \scriptfont2=\ninesy
      \scriptscriptfont2=\sixsy
    \textfont3=\twelveex          \scriptfont3=\ninex
      \scriptscriptfont3=\sixex
    \textfont\itfam=\twelveit    
    \textfont\slfam=\twelvesl    
      \scriptscriptfont\slfam=\sixsl
    \textfont\bffam=\twelvebf     \scriptfont\bffam=\ninebf
      \scriptscriptfont\bffam=\sixbf
    \textfont\ttfam=\twelvett
    \textfont\cpfam=\twelvecp
    \textfont\ssfam=\twelvess     \scriptfont\ssfam=\niness
      \scriptscriptfont\ssfam=\sixss
    \ifamsfonts
       \textfont\msafam=\twelvemsa     \scriptfont\msafam=\ninemsa
         \scriptscriptfont\msafam=\sixmsa
       \textfont\msbfam=\twelvemsb     \scriptfont\msbfam=\ninemsb
         \scriptscriptfont\msbfam=\sixmsb
       \textfont\eufam=\twelveeu     \scriptfont\eufam=\nineeu
         \scriptscriptfont\eufam=\sixeu \fi
    \samef@nt
    \b@gheight=12pt
    \setbox\strutbox=\hbox{\vrule height 0.85\b@gheight
                                depth 0.35\b@gheight width\z@ }}
\twelvepoint
%
%
\baselineskip = 15pt plus 0.2pt minus 0.1pt 
\lineskip = 1.5pt plus 0.1pt minus 0.1pt
\lineskiplimit = 1.5pt
\parskip = 6pt plus 2pt minus 1pt
\interlinepenalty=50
\interfootnotelinepenalty=5000
\predisplaypenalty=9000
\postdisplaypenalty=500
\hfuzz=1pt
\vfuzz=0.2pt
\dimen\footins=24 truecm 
\ifafour
 \hsize=16cm \vsize=22cm
\else
 \hsize=6.5in \vsize=9in
\fi
%
%
\skip\footins=\medskipamount
\newcount\fnotenumber
\def\clearfnotenumber{\fnotenumber=0} \clearfnotenumber
\def\fnote{\global\advance\fnotenumber by1 \generatefootsymbol
 \footnote{$^{\footsymbol}$}}
\def\fd@f#1 {\xdef\footsymbol{\mathchar"#1 }}
\def\generatefootsymbol{\iffrontpage\ifcase\fnotenumber
\or \fd@f 279 \or \fd@f 27A \or \fd@f 278 \or \fd@f 27B
\else  \fd@f 13F \fi
\else\xdef\footsymbol{\the\fnotenumber}\fi}
%
%
\newcount\secnumber \newcount\appnumber
\def\clearappnumber{\appnumber=64} \def\clearsecnumber{\secnumber=0}
\clearsecnumber \clearappnumber
\newif\ifs@c 
\newif\ifs@cd 
\s@cdtrue 
\def\unsectioned{\s@cdfalse\let\section=\subsection}
\newskip\sectionskip         \sectionskip=\medskipamount
\newskip\headskip            \headskip=8pt plus 3pt minus 3pt
\newdimen\sectionminspace    \sectionminspace=10pc
\def\Titlestyle#1{\par\begingroup \interlinepenalty=9999
     \leftskip=0.02\hsize plus 0.23\hsize minus 0.02\hsize
     \rightskip=\leftskip \parfillskip=0pt
     \advance\baselineskip by 0.5\baselineskip
     \hyphenpenalty=9000 \exhyphenpenalty=9000
     \tolerance=9999 \pretolerance=9000
     \spaceskip=0.333em \xspaceskip=0.5em
     \fourteenpoint
  \noindent #1\par\endgroup }
\def\titlestyle#1{\par\begingroup \interlinepenalty=9999
     \leftskip=0.02\hsize plus 0.23\hsize minus 0.02\hsize
     \rightskip=\leftskip \parfillskip=0pt
     \hyphenpenalty=9000 \exhyphenpenalty=9000
     \tolerance=9999 \pretolerance=9000
     \spaceskip=0.333em \xspaceskip=0.5em
     \fourteenpoint
   \noindent #1\par\endgroup }
\def\spacecheck#1{\dimen@=\pagegoal\advance\dimen@ by -\pagetotal
   \ifdim\dimen@<#1 \ifdim\dimen@>0pt \vfil\break \fi\fi}
\def\section#1{\cleareqnumber \s@ctrue \global\advance\secnumber by1
   \par \ifnum\the\lastpenalty=30000\else
   \penalty-200\vskip\sectionskip \spacecheck\sectionminspace\fi
   \noindent {\caps\enspace\S\the\secnumber\quad #1}\par
   \nobreak\vskip\headskip \penalty 30000 }
\def\undertext#1{\vtop{\hbox{#1}\kern 1pt \hrule}}
\def\subsection#1{\par
   \ifnum\the\lastpenalty=30000\else \penalty-100\smallskip
   \spacecheck\sectionminspace\fi
   \noindent\undertext{#1}\enspace \vadjust{\penalty5000}}

\def\appendix#1{\cleareqnumber \s@cfalse \global\advance\appnumber by1
   \par \ifnum\the\lastpenalty=30000\else
   \penalty-200\vskip\sectionskip \spacecheck\sectionminspace\fi
   \noindent {\caps\enspace Appendix \char\the\appnumber\quad #1}\par
   \nobreak\vskip\headskip \penalty 30000 }
\def\ack{\par\penalty-100\medskip \spacecheck\sectionminspace
   \line{\fourteencp\hfil ACKNOWLEDGEMENTS\hfil}%
\nobreak\vskip\headskip }
\def\refs{\begingroup \par\penalty-100\medskip \spacecheck\sectionminspace
   \line{\fourteencp\hfil REFERENCES\hfil}%
\nobreak\vskip\headskip \frenchspacing }
\def\endrefs{\par\endgroup}
%
%
\newif\iffrontpage \frontpagefalse
\headline={\hfil}
\footline={\iffrontpage\hfil\else \hss\twelverm
-- \folio\ --\hss \fi }
%
%
\newskip\frontpageskip \frontpageskip=12pt plus .5fil minus 2pt
\def\titlepage{\global\frontpagetrue\hrule height\z@ \relax
               \pubblock\relax }
\def\endtitlepage{\vfil\break\clearfnotenumber\frontpagefalse}
\def\title#1{\vskip\frontpageskip\Titlestyle{\caps #1}\vskip3\headskip}
\def\author#1{\vskip.5\frontpageskip\titlestyle{\caps #1}\nobreak}
\def\and{\par\kern 5pt \centerline{\sl and}}

\def\address#1{\par\kern 5pt\titlestyle{\it #1}}
\def\andaddress{\par\kern 5pt \centerline{\sl and} \address}

\def\abstract#1{\par\dimen@=\prevdepth \hrule height\z@ \prevdepth=\dimen@
   \vskip\frontpageskip\spacecheck\sectionminspace
   \centerline{\fourteencp ABSTRACT}\vskip\headskip
   {\noindent #1}}

\def\email#1{\fnote{\tentt e-mail: #1\hfill}}

%
%

%

%
\def\QMW{\address{%
   Department of Physics, Queen Mary and Westfield College\break
   Mile End Road, London E1 4NS, UK}}
%

%
%
\newcount\refnumber \def\clearrefnumber{\refnumber=0}  \clearrefnumber
\newwrite\R@fs                              
\immediate\openout\R@fs=\jobname.refs 
\def\closerefs{\immediate\closeout\R@fs} 
\def\refsout{\closerefs\refs
\unlockat
\input\jobname.refs
\lockat
\endrefs}
\def\refitem#1{\item{{\bf #1}}}
\def\ifundefined#1{\expandafter\ifx\csname#1\endcsname\relax}
\def\[#1]{\ifundefined{#1R@FNO}%
\global\advance\refnumber by1%
\expandafter\xdef\csname#1R@FNO\endcsname{[\the\refnumber]}%
\immediate\write\R@fs{\noexpand\refitem{\csname#1R@FNO\endcsname}%
\noexpand\csname#1R@F\endcsname}\fi{\bf \csname#1R@FNO\endcsname}}
\def\refdef[#1]#2{\expandafter\gdef\csname#1R@F\endcsname{{#2}}}
%
%
\newcount\eqnumber \def\cleareqnumber{\eqnumber=0}
\newif\ifal@gn \al@gnfalse  
\def\veqnalign#1{\al@gntrue \vbox{\eqalignno{#1}} \al@gnfalse}
\def\eqnalign#1{\al@gntrue \eqalignno{#1} \al@gnfalse}
\def\(#1){\relax%
\ifundefined{#1@Q}
 \global\advance\eqnumber by1
 \ifs@cd
  \ifs@c
   \expandafter\xdef\csname#1@Q\endcsname{{%
\noexpand\rm(\the\secnumber .\the\eqnumber)}}
  \else
   \expandafter\xdef\csname#1@Q\endcsname{{%
\noexpand\rm(\char\the\appnumber .\the\eqnumber)}}
  \fi
 \else
  \expandafter\xdef\csname#1@Q\endcsname{{\noexpand\rm(\the\eqnumber)}}
 \fi
 \ifal@gn
    & \csname#1@Q\endcsname
 \else
    \eqno \csname#1@Q\endcsname
 \fi
\else%
\csname#1@Q\endcsname\fi\global\let\@Q=\relax}
%
%
\newif\ifm@thstyle \m@thstylefalse
\def\mathstyle{\m@thstyletrue}
\def\proclaim#1#2\par{\smallbreak\begingroup
\advance\baselineskip by -0.25\baselineskip%
\advance\belowdisplayskip by -0.35\belowdisplayskip%
\advance\abovedisplayskip by -0.35\abovedisplayskip%
    \noindent{\caps#1.\enspace}{#2}\par\endgroup%
\smallbreak}
\def\m@kem@th<#1>#2#3{%
\ifm@thstyle \global\advance\eqnumber by1
 \ifs@cd
  \ifs@c
   \expandafter\xdef\csname#1\endcsname{{%
\noexpand #2\ \the\secnumber .\the\eqnumber}}
  \else
   \expandafter\xdef\csname#1\endcsname{{%
\noexpand #2\ \char\the\appnumber .\the\eqnumber}}
  \fi
 \else
  \expandafter\xdef\csname#1\endcsname{{\noexpand #2\ \the\eqnumber}}
 \fi
 \proclaim{\csname#1\endcsname}{#3}
\else
 \proclaim{#2}{#3}
\fi}
\def\Thm<#1>#2{\m@kem@th<#1M@TH>{Theorem}{\sl#2}}
\def\Prop<#1>#2{\m@kem@th<#1M@TH>{Proposition}{\sl#2}}
\def\Def<#1>#2{\m@kem@th<#1M@TH>{Definition}{\rm#2}}
\def\Lem<#1>#2{\m@kem@th<#1M@TH>{Lemma}{\sl#2}}
\def\Cor<#1>#2{\m@kem@th<#1M@TH>{Corollary}{\sl#2}}
\def\Conj<#1>#2{\m@kem@th<#1M@TH>{Conjecture}{\sl#2}}
\def\Rmk<#1>#2{\m@kem@th<#1M@TH>{Remark}{\rm#2}}
\def\Exm<#1>#2{\m@kem@th<#1M@TH>{Example}{\rm#2}}
\def\Qry<#1>#2{\m@kem@th<#1M@TH>{Query}{\it#2}}
%
%

%
\def\<#1>{\csname#1M@TH\endcsname}
%
%
\def\ref#1{{\bf [#1]}}
\def\ie{{\it i.e.\/}}
\def\nl{\hfil\break}
%
%

\def\lapprox{\hbox{\lower3pt\hbox{$\buildrel<\over\sim$}}}
\def\gapprox{\hbox{\lower3pt\hbox{$\buildrel<\over\sim$}}}
\def\quotient#1#2{#1/\lower0pt\hbox{${#2}$}}
\def\fr#1/#2{\mathord{\hbox{${#1}\over{#2}$}}}
\ifamsfonts
 \mathchardef\empty="0\hexmsb3F 
 \mathchardef\lsemidir="2\hexmsb6E 
 \mathchardef\rsemidir="2\hexmsb6F 
\else
 \let\empty=\emptyset
 \def\lsemidir{\mathbin{\hbox{\hskip2pt\vrule height 5.7pt depth -.3pt
    width .25pt\hskip-2pt$\times$}}}
 \def\rsemidir{\mathbin{\hbox{$\times$\hskip-2pt\vrule height 5.7pt
    depth -.3pt width .25pt\hskip2pt}}}
\fi
%
%

%
%
\def\integ{\mathord{\bb Z}} 
%
\def\str{\mathop{\rm str}}
%
\def\underrightarrow#1{\vtop{\ialign{##\crcr
      $\hfil\displaystyle{#1}\hfil$\crcr
      \noalign{\kern-\p@\nointerlineskip}
      \rightarrowfill\crcr}}} 
\def\underleftarrow#1{\vtop{\ialign{##\crcr
      $\hfil\displaystyle{#1}\hfil$\crcr
      \noalign{\kern-\p@\nointerlineskip}
      \leftarrowfill\crcr}}}  

\def\comm#1#2{\left[#1\, ,\,#2\right]}
%
\def\pder#1#2{{{\partial #1}\over{\partial #2}}}
%
%

\def\NPB#1#2#3{{\sl Nucl. Phys.} {\bf B#1} (#2) #3}

\def\CMP#1#2#3{{\sl Comm. Math. Phys.} {\bf #1} (#2) #3}

\def\PLA#1#2#3{{\sl Phys. Lett.} {\bf #1A} (#2) #3}
\def\PLB#1#2#3{{\sl Phys. Lett.} {\bf #1B} (#2) #3}
\def\JMP#1#2#3{{\sl J. Math. Phys.} {\bf #1} (#2) #3}

\def\AoP#1#2#3{{\sl Ann. of Phys.} {\bf #1} (#2) #3}

\def\RMP#1#2#3{{\sl Rev. Mod. Phys.} {\bf #1} (#2) #3}

\def\FAP#1#2#3{{\sl Funkt. Anal. Prilozheniya} {\bf #1} (#2) #3}
\def\FAaIA#1#2#3{{\sl Functional Analysis and Its Application} {\bf #1} (#2)
#3}

\def\Invm#1#2#3{{\sl Invent. math.} {\bf #1} (#2) #3}

\def\IJMPA#1#2#3{{\sl Int. J. Mod. Phys.} {\bf A#1} (#2) #3}

\def\AdM#1#2#3{{\sl Advances in Math.} {\bf #1} (#2) #3}

\def\TMP#1#2#3{{\sl Theor. Mat. Phys.} {\bf #1} (#2) #3}

\def\JSM#1#2#3{{\sl J. Soviet Math.} {\bf #1} (#2) #3}

\def\PJAS#1#2#3{{\sl Proc. Jpn. Acad. Sci.} {\bf #1} (#2) #3}
\def\JPSJ#1#2#3{{\sl J. Phys. Soc. Jpn.} {\bf #1} (#2) #3}
\def\JETPL#1#2#3{{\sl  Sov. Phys. JETP Lett.} {\bf #1} (#2) #3}
\def\JDG#1#2#3{{\sl J. Diff. Geometry} {\bf #1} (#2) #3}

\lockat

%
%

\let\pb=\anticomm

\def\spdo{{\hbox{S$\Psi$DO}}}

\def\fr#1/#2{\hbox{${#1}\over{#2}$}}
\def\Fr#1/#2{{{#1}\over{#2}}}

\def\sres{{\rm sres\,}}


\def\ope#1#2{{{#2}\over{\ifnum#1=1 {x-y} \else {(x-y)^{#1}}\fi}}}

\def\W{{\ssf W}}
\def\SBKP{{\ssf SBKP}}
\def\SDOP{{\ssf SDOP}}
\def\BKP{{\ssf BKP}}
\def\SKP{{\ssf SKP}}
\def\KP{{\ssf KP}}
\def\SKdV{{\ssf SKdV}}
\def\KdV{{\ssf KdV}}
\def\JSKP{{\ssf JSKP}}

\refdef[Adler]{M.~Adler, \Invm{50}{1981}{403}.}
\refdef[MaRa]{Yu.~I.~Manin and A.~O.~Radul, \CMP{98}{1985}{65}.}
\refdef[DickeyI]{L.~A.~Dickey, {\sl Lectures in field theoretical
Lagrange-Hamiltonian formalism}, (unpublished).}
\refdef[DickeyII]{L.~A.~Dickey, \CMP{87}{1982}{127}.}
\refdef[DickeyIII]{L.~A.~Dickey, {\sl Integrable equations and Hamiltonian
systems}, World Scientific.}
\refdef[GD]{I.~M.~Gel'fand and L.~A.~Dickey, {\sl A family of Hamiltonian
structures connected with integrable nonlinear differential equations},
Preprint 136, IPM AN SSSR, Moscow (1978).}
\refdef[DS]{V.~G.~Drinfel'd and V.~V.~Sokolov, \JSM{30}{1984}{1975}.}
\refdef[KW]{B.~A.~Kupershmidt and G.~Wilson, \Invm{62}{1981}{403}.}
\refdef[Mag]{F.~Magri, \JMP{19}{1978}{1156}.}
\refdef[Dickeypc]{L.~A.~Dickey, private communication.}
\refdef[STS]{M.~A.~Semenov-Tyan-Shanski\u\i, \FAaIA{17}{1983}{259}.}
\refdef[LM]{D.~R.~Lebedev and Yu.~I.~Manin, \FAP{13}{1979}{40}.}
\refdef[Uniw]{J.~M.~Figueroa-O'Farrill and E.~Ramos, {\sl Existence
and Uniqueness of the Universal $W$-Algebra}, to appear in
the {\sl J.~Math.~Phys.}}
\refdef[FL]{V.~A.~Fateev and S.~L.~Lykyanov, \IJMPA{3}{1988}{507}.}
\refdef[DFIZ]{P.~Di Francesco, C.~Itzykson, and J.-B.~Zuber,
\CMP{140}{1991}{543}.}
\refdef[KR]{V.~G.~Ka{\v c} and A.~C.~Raina, {\sl Lectures on highest
weight representations of infinite dimensional Lie algebras}, World
Scientific, etc...}
\refdef[KP]{E.~Date, M.~Jimbo, M.~Kashiwara, and T.~Miwa
\PJAS{57A}{1981}{387}; \JPSJ{50}{1981}{3866}.}
\refdef[DickeyKP]{L.~A.~Dickey, {\sl Annals of the New York Academy of
Science} {\bf 491} (1987) 131.}
\refdef[WKP]{J.~M.~Figueroa-O'Farrill, J.~Mas, and E.~Ramos,
\PLB{266}{1991}{298}.}
\refdef[Yama]{K. Yamagishi, \PLB{259}{1991}{436}.}
\refdef[YuWu]{F. Yu and Y.-S. Wu, {\sl Hamiltonian Structure,
(Anti-)Self-Adjoint Flows in KP Hierarchy and the $W_{1+\infty}$ and
$W_\infty$ Algebras}, Utah Preprint, January 1991.}
\refdef[ClassLim]{J. M. Figueroa-O'Farrill and E. Ramos, {\sl The
Classical Limit of $W$-Algebras}, \PLB{282}{1992}{357},({\tt hep-th/9202040}).}
\refdef[KoSt]{B. Kostant and S. Sternberg, \AoP{176}{1987}{49}.}
\refdef[TakaTake]{K. Takasaki and T. Takebe, {\sl SDIFF(2) KP
Hierarchy}, Preprint RIMS-814, December 1991.}
\refdef[KoGi]{Y. Kodama, \PLA{129}{1988}{223},
\PLA{147}{1990}{477};\nl
Y. Kodama and J. Gibbons, \PLA{135}{1989}{167}.}
\refdef[Bakas]{I. Bakas, \CMP{134}{1990}{487}.}
\refdef[Pope]{C. N. Pope, L. J. Romans, and X. Shen,
\PLB{242}{1990}{401}.}
\refdef[Lang]{S. Lang, {\sl Algebra}, Second Edition, Addison-Wesley
1984.}
\refdef[Radul]{A. O. Radul, \JETPL{50}{1989}{373}.}
\refdef[Morozov]{A. Morozov, \NPB{357}{1991}{619}.}
\refdef[LPSW]{H. Lu, C. N. Pope, X. Shen, and X. J. Wang, {\sl The
complete structure of $W_N$ from $W_\infty$ at $c=-2$}, Texas A\& M
Preprint CPT TAMU-33/91 (May 1991).}
\refdef[Winfty]{C. N. Pope, L. J. Romans, and X. Shen,
\PLB{236}{1990}{173},\PLB{242}{1990}{401},\NPB{339}{1990}{191};\nl
I. Bakas, \CMP{134}{1990}{487}.}
\refdef[Zam]{A. B. Zamolodchikov, \TMP{65}{1986}{1205}.}
\refdef[Douglas]{M. R. Douglas, \PLB{238}{1990}{176}.}
\refdef[Guill]{V. W. Guillemin, \AdM{10}{1985}{131}.}
\refdef[Wod]{M. Wodzicki,{\sl Noncommutative Residue} in {\sl K-Theory,
Arithmetic and Geometry}. Ed. Yu. I. Manin. Lectures Notes in
Mathematics 1289, Springer-Verlag.}
\refdef[Shubin]{M. A. Shubin, {\sl Pseudodifferential operators
and spectral Theory}, Springer-Verlag.}
\refdef[Class]{J. M. Figueroa-O'Farrill and E. Ramos, {\sl
Classical $W$-algebras from dispersionless Lax hierarchies
}, Preprint-KUL-TF-92/6, June 1992.}
\refdef[Mamoramos]{F.Mart{\'\i}nez-Mor{\'a}s and E. Ramos,
{\sl Higher Dimensional Classical W-algebras }, Preprint-KUL-TF-92/19 and
US-FT/6-92,({\tt hep-th9206040}).}
\refdef[Hull2]{C.M. Hull, {\sl W-geometry}, Preprint-QMW-92-6,
({\tt hep-th/9211113}).}
\refdef[Yu]{F. Yu,{\sl Bi-Hamiltonian Structure of Super KP
Hierarchy}, UU-HEP-91/13, ({\tt hep-th/9109009}).}
\refdef[N=1]{J.M. Figueroa-O'Farrill and E. Ramos,\CMP{145}{1992}{43}.}
\refdef[N=2]{J.M. Figueroa-O'Farrill and E. Ramos,\NPB{368}{1992}{361}.}
\refdef[Toda]{S. Komata, K. Mohri, and H. Nohara, \NPB{359}{1991}{168}.}
\refdef[virskdv]{P. Mathieu, \JMP{29}{1988}{2499}.}
\refdef[skpdos]{J.M. Figueroa-O'Farrill, J. Mas, and E. Ramos,
\RMP{3}{1991}{479}.}
\refdef[DargisMathieu]{P. Dargis and P. Mathieu, {\sl Nonlocal
conservation laws for supersymmetric KdV equations}, LAVAL-PHY -21/93,
({\tt hep-th/9301080}).}
\refdef[FiRa]{J.M. Figueroa-O'Farrill and E. Ramos,\PLB{262}{1991}{265}.}
\refdef[Oevel]{W. Oevel and Z. Popowicz, \CMP{139}{1991}{441}.}
\refdef[RamosStanciu]{E. Ramos and S. Stanciu, {\sl On the
Supersymmetric $\BKP$ Hierarchy}, QMW-PH-94-3, ({\tt hep-th/9402056}).}
\refdef[Kuper]{B.A. Kupershmidt, \PLA{102}{1984}{213}.}
\refdef[Kersten]{P.H.M. Kersten, \PLA{134}{1988}{25}.}
\refdef[Mulase]{M. Mulase, \JDG{34}{1991}{651}.}
\refdef[Rabin]{J. Rabin, \CMP{137}{1991}{533}.}
\refdef[Bergshoeff]{ E. Bergshoeff, C.N. Pope, L.J. Romans, and
X. Shen, \PLB{245}{1990}{447}.}
\refdef[Sonia]{S. Stanciu, {\sl Additional Symmetries of
Supersymmetric $\KP$ Hierarchies}, BONN-HE-93-31, ({\tt hep-th/9309058}).}
\refdef[Das]{A. Das, E. Sezgin, and J.Sin, \PLB{277}{1992}{435}.}
\unsectioned
\overfullrule=0pt

\def\pubblock{ \line{\hfil\twelverm Preprint-QMW-PH-94-4}
               \line{\hfil\twelverm  March 1994}
               \line{\hfil hep-th/9403043}}
\titlepage
\title{A comment on the odd flows for the
supersymmetric $\KdV$  equation.}
\author{Eduardo Ramos\email{ramos@v2.ph.qmw.ac.uk}}
\QMW
\abstract{ In a recent paper Dargis and Mathieu introduced
integrodifferential odd
flows for the supersymmetric $\KdV$ equation. These flows are
obtained from the nonlocal conservation laws associated with the
fourth root of its Lax operator. In this note I show that
only half of these flows are of the standard Lax form,
while the remaining half provide us with hamiltonians
for an $\SKdV$-type reduction of a new supersymmetric hierarchy.
This new hierarchy is shown to be closely
related to the Jacobian supersymmetric $\KP$-hierarchy of Mulase
and Rabin. A detailed study of the
algebra of additional symmetries of this new hierarchy reveals
that it is isomorphic to the super-$\W_{1+\infty}$ algebra, thus
making it a candidate for a possible interrelationship
between superintegrability and two-dimensional supergravity.
}

\endtitlepage

The history of the supersymmetric $\KdV$ equation, though short, has
already been full
of surprises. To the best of my knowledge, the first attempt to
supersymmetrize the $\KdV$ equation was carried out by Kuperschmidt
in \[Kuper]. He introduced an extra fermionic variable and was able
to define an extended system which was formally integrable
and bihamiltonian, but failed
to be invariant under supersymmetry. The first fully
supersymmetric $\KdV$ system---hereafter denoted by $\SKdV$---was
introduced by Manin and Radul in \[MaRa].
The $\SKdV$ equation is most naturally written in
$(1|1)$ superspace formalism\fnote{I will assume in what follows that
the reader is familiar with the standard superspace notation. A
concise reference, otherwise, is supplied by \[MaRa].} as
$${\partial U\over \partial t}= {1\over 4}U^{[6]} +
{3\over 4}(UU')''.\(skdvequation)$$
In full analogy with the bosonic case, the $\SKdV$ equation was
naturally obtained as a reduction of a supersymmetrization of
the $\KP$ hierarchy ($\SKP$). But unfortunately the analogies
stopped there. The Adler-Gel'fand-Dickey scheme which had proved
so fruitful in providing us with hamiltonian structures for $\KP$
and its reductions of the $\KdV$-type seemed, not to have an
analogue for $\SKP$.
Nevertheless, some time later,
Mathieu \[virskdv] showed that the
$\SKdV$ is hamiltonian with respect to the supervirasoro algebra, {\ie}
$$\pb{U(X)}{U(Y)}=\left({1\over 2}D^5 +{3\over 2} U(X) D^2
+{1\over 2} U'(X) D + U''(X)\right)\cdot\delta(X-Y).\(supervirasoro)$$
This made evident that the relationship between superintegrability and
supersymmetric Gel'fand-Dickey algebras was anything but lost.
This result notwithstanding, a bihamiltonian structure---a hallmark
of integrability---was still lacking.
Mathieu demonstrated that
a naive supersymmetrization of the first hamiltonian structure of
$\KdV$ failed to work and thus concluded that the $\SKdV$ equation
was not bihamiltonian.

Such was the state of affairs until Oevel and Popowizc \[Oevel],
and independently J.M. Figueroa-O'Farrill, J. Mas and the author
\[skpdos], observed
that the $\SKdV$ equation could be understood as a reduction of an
even-order $\SKP$-like operator, $\SKP_2$. This formalism had the
important advantage that a direct application of the supersymmetric
AGD scheme of \[FiRa] equipped the $\SKdV$ equation with a
bihamiltonian structure---the new hamiltonian structure being a
nonlocal deformation of the naive supersymmetrization of the first
hamiltonian structure of the $\KdV$ equation. Nevertheless, this approach
still suffered from two important drawbacks. First of all, the
reduction procedure was realized by an explicit computation using
Dirac brackets, which did not seem to be easily generalizable to
higher order $\SKdV$-type reductions. In fact, the emergence of the
supervirasoro algebra appeared unexpectedly due to ``miraculous
cancellations'' of explicit nonlocalities in the formalism. In
addition to this, the odd flows of the Manin-Radul $\SKP$ hierarchy
seemed to be irremediably lost for the $\SKP_2$ hierarchy.

The way out of this impasse was hinted at by Dargis and Mathieu in
\[DargisMathieu]. They realised that the first few nonlocal charges
for the $\SKdV$ equation obtained by Kersten \[Kersten] were given
by the fourth-order root of the $\SKdV$ Lax operator $D^4 +UD$.
In their paper, Dargis and Mathieu assumed without proof that
the Adler supertrace of $L^{{2k-1}\over 4}$
provided hamiltonians for Lax flows of the type
$$D_{2k-1}L =\comm{L}{(L^{{2k-1}\over 4})_-}.\(dargisflows)$$
But as I will explicitly show, only half of these flows, the ones
corresponding to $k$ an even integer, are consistent.
Moreover, the proof that these Lax flows are indeed hamiltonian
with respect the supervirasoro algebra, with hamiltonians
$$H_{{4k-1}\over 4}=-{4\over{4k-1}}\str L^{{4k-1}\over 4},\(oddtrace)$$
will be a simple exercise when using the machinery developed
in \[RamosStanciu].

This result opens the interesting question:  which
odd flows are obtained from the supertraces of $L^{{4k+1}\over 4}$
when we use the supervirasoro algebra as our Poisson structure?
The fact stated in \[DargisMathieu], that these hamiltonians were
not invariant under supersymmetry, seemed to suggest a possible
connection with the Mulase-Rabin Jacobian $\SKP$ hierarchy ($\JSKP$).
The main point of this note is to show that in fact this connection
exists, although the precise statement is that the resulting
odd flows are a linear combination of the odd flows of the $\SKP$ and
$\JSKP$ hierarchy. More interestingly, I will show that the algebra
of additional symmetries of this new hierarchy is nothing else but
the super-$\W_{1+\infty}$ algebra. Taking into account the relationship
in the bosonic case between additional symmetries of the $\KP$
equation and two-dimensional gravity, it does not seem outlandish to expect
a relationship between this new superintegrable hierarchy
and two-dimensional supergravity.

Understandably, lack of space will prevent me from giving a
detailed description of the formalism. Nonetheless, the reader
can find the required machinery in \[RamosStanciu] and references
therein.

\section{The Dargis-Mathieu odd flows}

The $\SKdV$ Lax operator $L=D^4 + UD$ can be nicely characterized as
the unique differential operator of order four obeying the
constraint
$$L^*=DLD^{-1}.\(skdvconst)$$
Compatibility of the odd flows \(dargisflows) with this constraint
requires
$$D_{2k-1}L^* =\comm{L}{(L^{{2k-1}\over 4})_-}^*,\(compatibility)$$
or equivalently
$$\comm{L}{(L^{{2k-1}\over 4})_-}^*= -D\comm{L}{(L^{{2k-1}\over 4})_-}
D^{-1},\(compatibilitydos)$$
which is only satisfied provided
$$ (D L^{{2k-1}\over 4} D^{-1})_- =
(-)^{k} D (L^{{2k-1}\over 4})_- D^{-1}.\(zeromode)$$
On the one hand, the leading term of the above equation clearly implies that
$k$ should be an even integer. On the other hand,
it is a simple computational matter to check that for an arbitrary
{\spdo}  $A=\sum_j A_j D^j$ the relation $(D A D^{-1})_- = D A_- D^{-1}$
holds if and only if $A_0 =0$, in other words if
$\sres L^{{2k-1}\over 4} D^{-1} =0$.
But this is automatically
fulfilled for $k$ an even integer. Indeed
$$\eqnalign{\sres L^{{2k-1}\over 4}
D^{-1} &=\sres (L^{{2k-1}\over 4} D^{-1})^*\cr
&=-\sres D^{-1} (L^*)^{{2k-1}\over 4}\cr
&=(-)^{k+1} \sres L^{{2k-1}\over 4} D^{-1}.\(zeroresidue)\cr}$$

In fact, a direct computation of the first flow with $k=1$ yields
$$(D_1 U)D = U' D +UU^{[-1]}+U'',\(example)$$
(where we use the notation $U^{[-j]}=(D^{-j}U)$),which is
clearly inconsistent because the left hand side is an operator with
no free term.

It was shown in \[RamosStanciu] how Poisson brackets for the $\SKdV$
variable could be induced from the natural ones associated to the Lax
operator $D^3 + U$. All the information about these Poisson brackets,
that in this particular case are nothing but the supervirasoro
algebra, can be neatly encoded in the Adler-type map
$$J (X) = (LX)_+L - LD^{-1}(DXL
D^{-1})_+ D,\(strucskdv)$$
where $X$ is a 1-form in the space of operators $L$, or
equivalently
$$X^*=(-)^{|X|+1}DXD^{-1}.\(simunaforma)$$
The gradients of the hamiltonians $H_{{4k-1}\over 4}$ can be directly
computed to yield
$$dH_{{4k-1}\over 4}= -L^{{2k-5}\over 4}.\(gradientsodd)$$
It is now a straightforward computation to check that
$$D_{4k-1}L= J(dH_{{4k-1}\over 4}),\(hamodd)$$
as conjectured by Dargis and Mathieu \[DargisMathieu].
\Rmk<>{ I have been rather cavalier with the use of the formalism
due to the fact that the hamiltonians $H_{{4k-1}\over 4}$ are not
differential polynomials in $U$, but rather integro-differential
polynomials.
Nevertheless the reader can check that all the necessary
manipulations to arrive to \(hamodd) can be extended to the
integrodifferential case.}

In order to understand what is going on with the flows induced by
the other half of the hamiltonians it will be necessary to make
a small disgression about $\SKdV$-type reductions of the $\JSKP$
hierarchy.

\section{The Jacobian $\SKP$ hierarchy}

It is well known that the $\SKP$ hierarchy of Manin and Radul
and $\SKP_2$ are not
the only possible extension to superspace of the standard $\KP$ hierarchy.
In references \[Mulase] and \[Rabin] Mulase and Rabin proposed a
different supersymmetric extension of $\KP$. Their approach was
inspired by the geometrical interpretation of the standard $\KP$ equations
as flows in a Jacobian variety. From the purely algebraic point
of view the Jacobian $\SKP$ ($\JSKP$) flows are given as flows in the
superVolterra group. An element $\phi$ of the superVolterra group is
given by \spdo's of the form
$$\phi = 1 +\sum_{i=1}^{\infty} s_i D^{-i}.\(superVolterra)$$
The explicit expression of the $\JSKP$ flows is then given by
$$\eqnalign{
D_{2k}\phi =& -\left(\phi\partial^k\phi^{-1}\right)_-\phi\(JSKPuno)\cr
D_{2k-1}\phi =&-\left(\phi\partial^{k-1}\partial_{\theta}
\phi^{-1}\right)_-\phi.\(JSKPdos)\cr}$$

It is now simple to check that the even flows of $\JSKP$ are
equivalent to the ones of $\SKP$ and $\SKP_2$ under the assumption
that the Lax operators defining this two hierarchies are dressable.
In general, for a dressable {\spdo} $\Lambda$ of order $j$, take $\phi$ such
that
$$\Lambda=\phi D^j \phi^{-1}.\(dressing)$$
Then
$$D_{2k}\Lambda =\comm{\Lambda}{\Lambda^k_-}\quad\Leftrightarrow
D_{2k}\phi = -\left(\phi\partial^k\phi^{-1}\right)_-\phi.\()$$
But in contrast the odd flows induced on $\Lambda$ through \(JSKPdos) are
given by
$$D_{2k-1}\Lambda=\comm{\Lambda}{(\Lambda^{k-1}M)_-},\(oddfunny)$$
with $M=\phi\partial_{\theta}\phi^{-1}$. Since $M$ cannot be simply
expressed in terms of $\Lambda$, it
is costumary to say that the odd flows of $\JSKP$ are
not of the Lax form, although in fact they are.

The obvious question to ask oneself
now is: what happens with the $\JSKP$ flows when
a reduction of the $\SKdV$-type is imposed upon them? Of course there
is nothing new to say about the even flows because they are the ``good
old ones'', but what about the odd ones?

Let us first study what is the effect of the $\SKdV$ constraints at
the level of the dressing field $\phi$. Because of the results of
\[RamosStanciu] we can restrict ourselves without lost of generality
to the $\SKP_2$ case, {\ie} $\Lambda = D^2 +\cdots$.
As before we should first consider the $\SBKP$
constraint $\Lambda^*=-D\Lambda D^{-1}$. At the level of the $\phi$
this implies that
$$(\phi^{-1})^*\partial\phi^* = D\phi\partial\phi^{-1}D^{-1}.\(choruno)$$
If we now use that $(\phi^{-1})^* = (\phi^*)^{-1}$,  {\(choruno)} can be
written as $\partial = P\partial P^{-1}$, with
$P=\phi^*D\phi D^{-1}$, thus
$$\phi^* =D\phi^{-1}D^{-1}.$$
Thereby,
$$\eqalign{
M^* =-& (\phi^{-1})^*\partial_{\theta}\phi^*\cr
=& -D\phi D^{-1}\partial_{\theta} D\phi^{-1}D^{-1}\cr
=& D\phi\partial_{\theta}\phi^{-1}D^{-1} -
D\phi D^{-1}\comm{\partial_{\theta}}{D}\phi^{-1}D^{-1}\cr
=&D\left(\phi \partial_{\theta}\phi^{-1} -\phi D
\phi^{-1}\right)D^{-1}\cr
=& D M D^{-1} - D \Lambda^{1\over 2}D^{-1}.\cr
}$$
And consequently the odd flows do not reduce nicely. But fortunately
not everything is lost.
Notice that the combination $S= 2M - \Lambda^{1\over 2}$ has the
``correct'' symmetry properties, {\ie}
$$S^* = DSD^{-1},\()$$
and moreover $S^2 =\Lambda$.
Nevertheless, I will keep the notation $\Lambda^{1\over 2}$ for the
manifestly supersymmetric square root of $\Lambda$, {\ie},
the dressed version of the superderivative.
Curiously enough, $S$ is nothing but $\phi Q\phi^{-1}$, where $Q$
is the generator of supersymmetry transformations.

As a result of all of this, it seems natural to introduce the following
$\JSKP$-like hierarchy with oddflows given by
$$D_{2k-1}\Lambda=\comm{\Lambda}{(\Lambda^{k-1}S)_-},\(newflows)$$
which can be understood as a linear combination of the original
$\SKP$ and $\JSKP$ odd flows. It is now simple to show that these
flows are indeed consistent with the $\SBKP$ constraint as long
as $k=2j +1$.
As before, the consistency of the oddflows requires
$$\sres \Lambda^{2j}S D^{-1} =0,\()$$
and this follows from
$$\eqalign{
\sres \Lambda^{2j}S D^{-1} =&\sres \left( \Lambda^{2j}S D^{-1}\right)^*\cr
=&-\sres D^{-1}S^*(\Lambda^{2j})^*\cr
=&-\sres \Lambda^{2j} S D^{-1}.\cr}$$

\Rmk<>{When there is an explicit dependence on $\partial_{\theta}$, in
order to compute the super-residue, or equivalently the $+$ and
$-$ projections, we should use that
$$\partial_{\theta} = D -\theta\partial,$$
and then apply the standard definitions \[MaRa].}

Because $D_{4j+1}$ acts as a superderivation it follows
that
$$D_{4j+1}\Lambda^n=\comm{\Lambda^n}{(\Lambda^{2j}S)_-},\(superder)$$
and it is consistent to impose the constraint $(\Lambda^n)_-=0$. In
particular for $n=2$ we recover the standard $\SKdV$ Lax operator.

I will give strong evidence in the following that for $n=2$ the flows
\(superder) are hamiltonian with respect the superVirasoro algebra,
and hamiltonians
$$H_{4j+1\over 4} \sim \str L^{{4j+1}\over 4}.\()$$
Moreover, the formalism is such
that the generalization of this result to higher order $\SKdV$-type
reductions is self-evident.

The key point is supplied by the relationship
$$dH_{{4j+1}\over 4} \sim L^{j-1} S.\(keypoint)$$
Notice first that as given above $dH_{{4j+1}\over 4}$ has the correct
weight and symmetry properties, {\ie}
$$dH_{{4j+1}\over 4}^*= DdH_{{4j+1}\over 4}D^{-1},$$
as corresponds to an odd one-form \(simunaforma).
It is now straightforward to show that
$$D_{4j+1} L \sim J(dH_{{4j+1}\over 4}).\()$$
Unfortunately, I do not have at this point a  proof of \(keypoint),
although I have checked its validity for the first few flows.

Notice that the first flow of the hierarchy is given by
$$D_1 U =\comm{Q}{U}= U' - \theta U'',\(firstflow)$$
which is nothing but the supersymmetric variation of $U$. This
property of $D_1$ is not a peculiarity of the $\SKdV$ reduction and
is also shared by the whole hierarchy. This is in
contrast to the $\SKP$ and $\JSKP$ hierarchies where the first flow
contains an explicit dependence on the potentials.

\section{Additional symmetries}

It is now simple to show, following standard techniques \[DickeyIII]
\[Das]\[Sonia], that the algebra
of additional symmetries of the new supersymmetric hierarchy given
by the standard even flows plus \(newflows) is
isomorphic to the algebra of superdifferential operators in $(1|1)$
superspace ($\SDOP$), which is itself
isomorphic to super-$\W_{1 +\infty}$ \[Bergshoeff].

The algebra of the flows is given by
$$\comm{D_{2i}}{D_{2j}}=0,\quad
\comm{D_{2i}}{D_{2j-1}}=0,\quad {\rm and}\
\comm{D_{2i-1}}{D_{2j-1}}= 2 D_{2i + 2j -2}.$$
A representation of these flows in terms of even and odd
time parameters is given by
$$\eqnalign{D_{2k}=&\pder{\ }{t_{2k}}\(flujopar)\cr
D_{2k-1}=&\pder{\ }{\tau _{2k-1}}
+ \sum_{j\geq i}\tau_{2j-1}\pder{\ }{t_{2k+2j-2}}\(flujoimpar)\cr}$$

In complete analogy  with the bosonic case the additional symmetries
are going to be generated by additional flows of the form
$$\partial_{\Gamma}\phi = -(\phi\Gamma\phi^{-1})_-\phi,\()$$
with
$$\eqnalign{
\comm{D_{2k}-\partial^{k}}{\Gamma}&=0,\(conduno)\cr
\comm{D_{2k-1}-\partial^{k-1}Q}{\Gamma}&=0.\(condos)\cr}$$
An obvious solution for $\Gamma$ is given by
the superderivative $D$. It is also possible to obtain
two other solutions by the standard procedure of deforming the
operators $x$ and $\theta$, {\ie}
$$\eqnalign{
\Gamma_x =& x + \sum_{k\geq 1} k t_{2k}\partial^{k-1}
+\sum_{k\geq 1} (k-1)\tau_{2k-1}\partial^{k-2}Q +\cr
&\theta \sum_{k\geq 1} \tau_{2k-1}\partial^{k-1}-
{1\over 2}\sum_{k,j\geq 1}(k-j)\tau_{2k-1}\tau_{2j-1}\partial^{i+j-2},
\()\cr}$$
and
$$\Gamma_{\theta}= \theta + \sum_{k\geq 1}\tau_{2k-1}
\partial^{k-1}.\()$$

Therefore any $\Gamma$ of the form
$$\Gamma = \Gamma_x^n\Gamma_{\theta}^k D^m,\()$$
with $n\geq 0$, $k=0,1$, and $m\in \integ$ defines a symmetry of
the hierarchy. Notice that the commutation relations
$$\comm{D}{\Gamma_x}=\Gamma_\theta,\ \ {\rm and}\ \ \comm{D}
{\Gamma_{\theta}}=1,\()$$
directly imply that the algebra of additional symmetries is isomorphic
to $\SDOP$.
\ack
I would like to thank J.M. Figueroa-O'Farrill, J. Mas,
J. Petersen, A.M.
Semikhatov, and S. Stanciu for many useful conversations on the
subject.

\refsout

\bye